\documentclass[twocolumn,english,showabstract, groupaddress, showkeys, showpacs]{revtex4}
\usepackage[T1]{fontenc}
\usepackage[latin9]{inputenc}
\usepackage{color}
\usepackage{units}
\usepackage{textcomp}
\usepackage{amstext}
\usepackage{graphicx}
\usepackage{esint}

\makeatletter

\providecommand{\tabularnewline}{\\}

\@ifundefined{textcolor}{}
{%
 \definecolor{BLACK}{gray}{0}
 \definecolor{WHITE}{gray}{1}
 \definecolor{RED}{rgb}{1,0,0}
 \definecolor{GREEN}{rgb}{0,1,0}
 \definecolor{BLUE}{rgb}{0,0,1}
 \definecolor{CYAN}{cmyk}{1,0,0,0}
 \definecolor{MAGENTA}{cmyk}{0,1,0,0}
 \definecolor{YELLOW}{cmyk}{0,0,1,0}
 }

\makeatother

\usepackage{babel}

\begin{document}

\title{Magnetocrystalline anisotropy in RAu$_{2}$Ge$_{2}$ (R = La, Ce
and Pr) single crystals}

\author{Devang A. Joshi}
\email{devang@tifr.res.in}
\author{A.~K. Nigam, S.~K. Dhar and A. Thamizhavel}
\affiliation{Department of Condensed Matter Physics and Materials Science,
Tata Institute of Fundamental Research, Colaba, Mumbai 400 005, India.}


\begin{abstract}
Anisotropic magnetic properties of single crystalline RAu$_{2}$Ge$_{2}$
(R = La, Ce and Pr) compounds are reported. LaAu$_{2}$Ge$_{2}$ exhibit
a Pauli-paramagnetic behavior whereas CeAu$_{2}$Ge$_{2}$ and PrAu$_{2}$Ge$_{2}$
show an antiferromagnetic ordering with N$\grave{e}$el temperatures
T$_{N}$ = 13.5 and 9~K, respectively. The anisotropic
magnetic response of Ce and Pr compounds establishes {[}001{]} as
the easy axis of magnetization and a sharp spin-flip type metamagnetic
transition is observed in the magnetic isotherms. The
resistance and magnetoresistance behavior of these compounds, in particular
LaAu$_{2}$Ge$_{2}$, indicate an anisotropic Fermi surface. The magnetoresistivity
of CeAu$_{2}$Ge$_{2}$ apparently reveals the
presence of a residual~Kondo interaction. A crystal
electric field analysis of the anisotropic susceptibility in conjunction
with the experimentally inferred Schottky heat capacity enables us
to propose a crystal electric field level scheme for Ce and Pr compounds.
For CeAu$_{2}$Ge$_{2}$ our values are in excellent agreement with
the previous reports on neutron diffraction. The
heat capacity data in LaAu$_{2}$Ge$_{2}$ show clearly the existence
of Einstein contribution to the heat capacity.
\end{abstract}



\maketitle

\section{Introduction}

The family of rare earth intermetallic compounds RT$_{2}$X$_{2}$
(R: rare earths, T: Transition metal and X: p-block element (Si or
Ge)) crystallizing in the ThCr$_{2}$Si$_{2}$-type tetragonal structure
constitute a reservoir of compounds showing a wide variety of interesting
magnetic and superconducting properties. Some of the Ce and Yb based
compounds within this family are well-known for their heavy fermion,
valence fluctuation, quantum criticality, superconductivity and~Kondo
behavior arising from the hybridization between the 4\textit{f} and
the conduction electrons. For example CeCu$_{2}$Si$_{2}$~\cite{Franz}is
a well~Known first heavy fermion superconductor, CeCu$_{2}$Ge$_{2}$
shows~Kondo behavior~\cite{Knopp}, CeRu$_{2}$Si$_{2}$~\cite{Yoshida},
CePd$_{2}$Si$_{2}$~\cite{Mathur}, CeCu$_{2}$Ge$_{2}$~\cite{Vargoz}
and CeRh$_{2}$Si$_{2}$\cite{Movshovich} show a quantum critical
behavior. \textcolor{black}{YbRh$_{2}$Si$_{2}$ possessing a low
lying antiferromagnetic ground state has been intensively studied
in recent years for its quantum critical behavior induced both by
external magnetic field and pressure.} Different types of magnetic
ordering are observed in the Pr compounds. PrRu$_{2}$Si$_{2}$~\cite{Mulders},
PrRu$_{2}$Ge$_{2}$~\cite{Vejpravova} and PrOs$_{2}$Si$_{2}$~\cite{Hiebl}
are ferromagnetic while antiferromagnetic or complicated magnetic
ordering is observed in case of PrCo$_{2}$Ge$_{2}$~\cite{Szytula},
PrCo$_{2}$Si$_{2}$~\cite{Kawae}, PrNi$_{2}$Si$_{2}$~\cite{Blanco}
and PrCu$_{2}$Ge$_{2}$~\cite{Szytula-1}.While
initial reports are typically based on polycrystalline materials,
data obtained on single crystals provide a more comprehensive picture
of the physical properties including their anisotropy, inevitable
in tetragonal symmetry. For example, our recent work on a single crystal
of CeAg$_{2}$Ge$_{2}$~\cite{Thamizh} removed the ambiguity about
the magnetic transition temperature of this compound which existed
due to the conflicting reports in the literature based on polycrystalline
samples. A quasi-quartet ground state arising due to two low lying
crystal electric field split states was also established from the
analysis of the magnetization data which show a large anisotropy along
the two principle crystallographic directions. The successful growth
of single crystals of CeAg$_{2}$Ge$_{2}$~\cite{Thamizh} achieved
for the first time motivated us to investigate the single crystalline
behavior of RAu$_{2}$Ge$_{2}$ ( R = La, Ce and Pr) compounds using
magnetic, thermal and transport measurements down to 1.8~K.

In this paper, findings of the detailed study of the physical properties
of single crystalline RAu$_{2}$Ge$_{2}$ (R = La, Ce and Pr) compounds
is reported. Polycrystalline CeAu$_{2}$Ge$_{2}$ and PrAu$_{2}$Ge$_{2}$
have previously been reported to order antiferromagnetically at \textcolor{black}{T$_{N}$
= 16~K}~\cite{Loidl} and 12~K~\cite{Nishimura} respectively. The
neutron diffraction results on CeAu$_{2}$Ge$_{2}$ clarifies the
pure antiferromagnetic ordering of the Ce moment lying along the \textit{c}
axis of the tetragonal unit cell with an ordered moment of 1.88~\textmu{}$_{B}$/Ce at 1.5~K. PrAu$_{2}$Ge$_{2}$ undergoes a metamagnetic transition at 32~KOe~\cite{Nishimura}. There are no reports on LaAu$_{2}$Ge$_{2}$. Our study on single crystalline samples infers the
ordering temperatures of the compounds to be 13.5 and 9~K for CeAu$_{2}$Ge$_{2}$
and PrAu$_{2}$Ge$_{2}$ respectively in contrast to the polycrystalline
report. Other results on CeAu$_{2}$Ge$_{2}$are in agreement with
the reported ones. In case of PrAu$_{2}$Ge$_{2}$ spin flip type
metamagnetic transition occurs at 22~KOe in contrast to the polycrystalline
report.

\section{Experimental}

Single crystals of RAu$_{2}$Ge$_{2}$ compounds were grown by high
temperature solution growth method. Since all the constituents of
RAu$_{2}$Ge$_{2}$ have high melting points, none of them could be
used as a flux. Looking into the binary phase diagram of the Au:Ge
system, it was observed that for a particular composition of Au:Ge
= 72:28, an eutectic forms with a melting point of 361~$^{\mathrm{o}}$C.
Since we had previously succeeded in growing the
single crystals of CeAg$_{2}$Ge$_{2}$ by using a binary Ag-Ge eutectic
as a self flux, initially we attempted Au-Ge eutectic for the growth
of RAu$_{2}$Ge$_{2}$ compounds. But the single crystals thus obtained
were small in size. We also attempted to grow the single crystals
of RAu$_{2}$Ge$_{2}$ compounds using Bi as flux. To avoid the reaction
of flux with the constituents of compounds, an ingot of RAu$_{2}$Ge$_{2}$
was prepared by melting in an arc furnace with the constituents taken
in proper stoichiometric ratio. The \textcolor{black}{as-cast ingot
of RAu$_{2}$Ge$_{2}$ and the Bi flux in the ratio of 1:19 were placed
in an alumina crucible and subsequently sealed in an evacuated quartz
tube. The mixture was heated up to 1100$^{\mathrm{o}}$C and maintained
at that temperature for 30 hrs}, followed by slow cooling (1$^{\mathrm{o}}$C
/ h) to 550$^{\mathrm{o}}$C and then rapidly cooled down to room
temperature. The crystals were separated from the flux by means of
centrifuging. The typical size of the obtained crystals was 7 $\times$
4 $\times$ 1 mm$^{3}$. The crystals nucleated on the walls of crucible
and formed as thin platelets sticking together to form a big single
crystal. An energy dispersive x-ray analysis EDAX was performed on
all of the obtained single crystals to identify their phase. The EDAX
results confirmed the crystals to be of the composition 1:2:2. In
our present study we have used the single crystals grown from the
Bi flux. To check for the phase purity, powder x-ray diffraction pattern
of these compounds were recorded by powdering a few small pieces of
single crystals. The grown crystals were then oriented along the principal
crystallographic directions by means of Laue diffraction. Well defined
Laue diffraction spots, together with the tetragonal symmetry pattern,
indicated the good quality of single crystals. The crystals were then
cut along the crystallographic directions using a spark erosion cutting
machine to study their anisotropic properties. The DC magnetic measurements
were performed in the temperature range 1.8 - 300~K and in the magnetic
fields up to 120~KOe along the two principal directions using a
SQUID magnetometer (Quantum Design) and a vibrating sample magnetometer
(VSM, Oxford Instruments). The resistivity and heat capacity were
measured using a physical property measurement system (PPMS, Quantum
Design).

\section{Results and Discussion}

\begin{figure}
\includegraphics[width=0.5\textwidth]{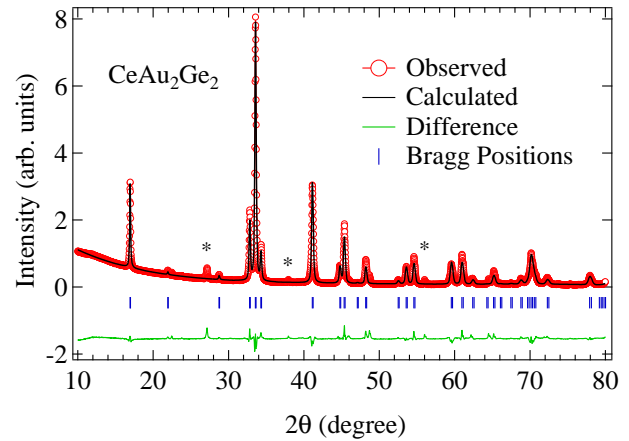}

\caption{\label{Fig. Rietveld}(Color online) Powder x-ray diffraction pattern
recorded for crushed single crystals of CeAu$_{2}$Ge$_{2}$ at room
temperature. The solid line through the experimental data points is
the Rietveld refi{}nement profi{}le calculated for the tetragonal
CeAu$_{2}$Ge$_{2}$. The stars represent the x-ray peaks corresponding
to Bi. }

\end{figure}
The RAu$_{2}$Ge$_{2}$ (R = La, Ce and Pr ) compounds form in a tetragonal
structure with a space group \textit{I4/mmm}. To confi{}rm the phase
homogeneity of the compounds with proper lattice and crystallographic
parameters, a Rietveld analysis of the observed x-ray pattern of the
three compounds was carried out using the FULLPROF program~\cite{Rodrigues}.
The lattice parameters and the unit cell volume thus obtained are
listed in Table I and a representative Rietveld refi{}ned plot of
CeAu$_{2}$Ge$_{2}$ is shown in Fig.~\ref{Fig. Rietveld}. We obtained
a $\chi^{2}$ value of 2.6, goodness of fit of 1.7, and Bragg R
factor of 0.081.%
\begin{table}
\begin{tabular}{ccccc}
 &  &  &  & \tabularnewline
\hline
\hline 
\hspace{1mm}Compound \hspace{1mm} & \hspace{1mm}\textit{a} ($\textrm{\AA}$) \hspace{1mm} & \hspace{1mm}\textit{c} ($\textrm{\AA}$) \hspace{1mm} & \hspace{1mm}V ($\textrm{\AA}^{3}$) \hspace{1mm} & \hspace{1mm}T$_{N}$ (K)\hspace{1mm}\tabularnewline
\hline
LaAu$_{2}$Ge$_{2}$ & 4.422 & 10.45 & 204.3 & P-P\tabularnewline
CeAu$_{2}$Ge$_{2}$ & 4.385 & 10.444 & 200.8 & 13.5\tabularnewline
PrAu$_{2}$Ge$_{2}$ & 4.366 & 10.443 & 199 & 9.0\tabularnewline
\hline
\hline 
 &  &  &  & \tabularnewline
\end{tabular}\caption{Lattice parameters of RAu$_{2}$Ge$_{2}$ compounds with unit cell
volume and N$\grave{e}$el temperature. P-P: Pauli paramagnetic.}

\end{table}
 The lattice parameters are comparable to those reported earlier for
the polycrystalline samples~\cite{Loidl,Nishimura}. The lattice parameters
decreases as we move from La to Pr, attributed to well~Known lanthanide
contraction. The unit cell volume of RAu$_{2}$Ge$_{2}$ compounds
is higher compared to that of RCu$_{2}$Ge$_{2}$ compounds and less
than that of RAg$_{2}$Ge$_{2}$ compounds. This may be due to the
intermediate size of the Au atom compared to Cu and Ag.

\subsection{LaAu$_{2}$Ge$_{2}$}

\begin{figure}
\includegraphics[width=0.5\textwidth]{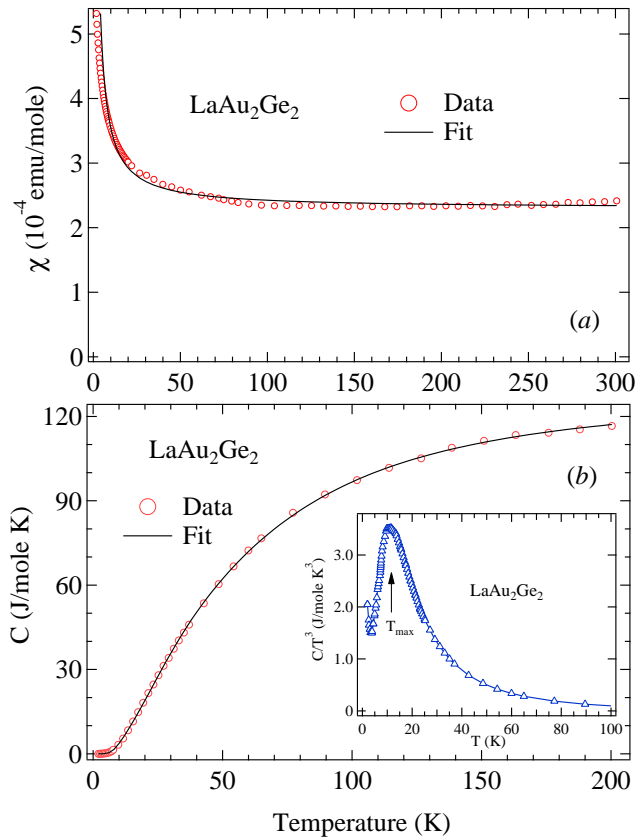}\caption{(Color online)\label{Fig. MT-HC_La} a) Magnetic susceptibility of
LaAu$_{2}$Ge$_{2}$ with a fit described in text. The inset shows
the C/T \textit{vs} T$^{2}$ plot with a linear fit. b) Heat capacity
curve of LaAu$_{2}$Ge$_{2}$ with a fit described in the text. The
inset shows the C/T$^{3}$\textit{vs} T plot.}

\end{figure}

We first describe the physical properties of LaAu$_{2}$Ge$_{2}$
which can be considered as the reference, non-magnetic analog for
the magnetic RAu$_{2}$Ge$_{2}$ compounds. The susceptibility of
LaAu$_{2}$Ge$_{2}$ (Fig.~\ref{Fig. MT-HC_La}a) shows a Pauli-paramagnetic
behavior at room temperature, with an absolute value of nearly 2.3$\times$10$^{-4}$~emu/mol. It remains nearly temperature independent down to 50~K and shows an upturn at lower temperatures (Fig.~\ref{Fig. MT-HC_La}a).
The low temperature upturn is most likely due to the presence of paramagnetic
ions in the constituents used to prepare the alloys. A fit (shown
by the solid line in Fig.~\ref{Fig. MT-HC_La}a) of the modified Curie-Weiss
law\begin{equation}
\chi=\chi_{0}+\frac{N\mu_{eff}^{2}}{3k_{B}(T-\theta_{p})}\label{Eq. modified Curie-Weiss}\end{equation}

to the data,\textcolor{blue}{{} }\ where the parameters have their
usual meaning, yields $\chi_{0}$ = 2.312 x 10$^{-4}$~emu/mol and
$\mu_{eff}$= 0.1~$\mu_{B}$. The magnitude of $\chi_{0}$ is typical
for the La compounds and a low value of the effective moment indicates
that the upturn at low temperatures is due to the presence of some
(< 1 \%) magnetic impurity ions in the constituents. The heat capacity
of the compound (Fig.~\ref{Fig. MT-HC_La}b) increases monotonically
with temperature. The magnitude of the electronic
contribution $\gamma$, obtained
from the low temperature heat capacity data is $\approx$7~mJ/mol
K$^{2}$. Inset of Fig. 2b shows a plot of C/T$^{3}$
\textit{vs} T , a representation
that is often used to assess the possible presence of low-frequency
Einstein modes in the specifi{}c heat~\cite{Lawless}. The maximum,
in this case at T$_{max}$ = 13~K, could be interpreted as the temperature
below which the Einstein modes are frozen out. This is also the temperature
where the deviation from a pure Debye description of the specific
heat becomes conspicuous. Considering this, the thermal variation
of the heat capacity of the compound was fitted to the combined Einstein
and Debye contributions as shown by the solid line in Fig.~\ref{Fig. MT-HC_La}b.
The total heat capacity in such a case is given by\begin{equation}
C_{Tot}=\gamma T+(C_{E}+C_{D})\label{Eq. Total-HC}\end{equation}
where the first term represents \ the electronic
contribution, the second term includes Einstein contribution C$_{E}$
and Debye contribution C$_{D}$. The Einstein contribution is given
by \begin{equation}
C_{E}=\sum_{n'}3n_{En'}R\frac{y^{2}e^{y}}{\left(e^{y}-1\right)^{2}}\label{Eq. Einstein}\end{equation}
where $y$ = $\Theta_{En'}/T$, $\Theta_{E}$ is the Einstein temperature,
$n'$ is the summation over the different Einstein temperatures, $R$
is the gas constant and $n_{E}$ is the number of Einstein oscillators. The Debye contribution is given by

\textcolor{black}{\begin{equation}
C_{D}=9n_{D}R\left(\frac{T}{\Theta_{D}}\right)^{3}\intop_{0}^{\Theta_{D}/T}\frac{x^{4}e^{x}dx}{\left(e^{x}-1\right)^{2}}\label{Eq. Debye}\end{equation}
where} $x$= $\Theta_{D}/T$. $\Theta_{D}$ is the Debye temperature
and $n_{D}$ is the number of Debye oscillators. Iterative fit to
the Eq.~\ref{Eq. Total-HC} was performed by using the values of electronic
contribution $\gamma$ as estimated above and fixing the number of
atoms $n_{D}$ and $n_{E}$ for a particular fit, allowing both $\Theta_{En'}$
and $\Theta_{D}$ to vary as fitting parameters. A good fit to the
heat capacity of LaAu$_{2}$Ge$_{2}$ over the entire range of temperature
was obtained by assigning three Debye characteristic atoms ($n_{D}$
= 3) with $\Theta_{D}=\mathrm{305}$~K plus two Einstein characteristic
atoms ($n_{E1}=n_{E2}=1$) with $\Theta_{E1}=74$~K . The description
of heat capacity in terms of a combination of acoustic and optical
modes can be readily understood by assigning the La and Ge atoms in
the unit cell to three Debye characteristic modes and Au to the remaining
Einstein modes. Since Au is much larger in size compared to other
atoms, it can be expected to vibrate with a lower
natural frequency. It can be shown that the Einstein temperature $\Theta_{E}$
is related to the peak in C/T$^{3}$ against T plot (T$_{max}$) by
the relation T$_{max}$ = $\Theta_{E}/5$~\cite{Chambers}, which
is in fair agreement with the observed value of T$_{max}$ = 13~K.

\begin{figure}
\includegraphics[width=0.5\textwidth]{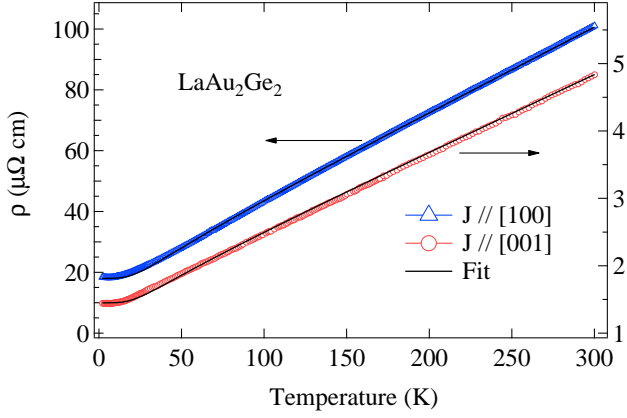}\caption{(Color online)\label{Fig. Res_La} Resistivity of LaAu$_{2}$Ge$_{2}$
with current parallel to {[}100{]} and {[}001{]} directions. The resistivity
along the two directions is fitted to Bloch-Gr$\ddot{\mathrm{u}}$nneisen
relation.}

\end{figure}
The resistivity of LaAu$_{2}$Ge$_{2}$ with current parallel to {[}100{]}
and {[}001{]} directions is shown in Fig.~\ref{Fig. Res_La}. The
resistivity along both the crystallographic directions
exhibit a metallic behavior down to 15~K and levels off at low temperatures
with a residual resistivity of 18~$\mu\Omega\mathrm{\, cm}$ and 1.4~$\mu\Omega\mathrm{\, cm}$ respectively for {[}100{]} and {[}001{]}
directions. The observed behavior is in tune with the phonon induced
scattering of the charge carriers as expected for a non magnetic compound.
The resistivity along {[}001{]} direction is found
to be lower compared to the in-plane {[}100{]} resistivity by a factor
of nearly 20 at 300~K points out significant anisotropy. The
similar anisotropic behavior in the resistivity was also found for
Ce and Pr compounds described below and may arise due to the inherent
structural anisotropy of the compound. The resistivity with current
parallel to the two crystallographic directions was fitted to the
modified Bloch-Gr$\mathrm{\ddot{u}}$neisen relation given by \begin{equation}
\rho(T)=\rho_{0}+4\Theta_{D}R\left(\frac{T}{\Theta_{D}}\right)^{5}\intop_{0}^{\Theta_{D}/T}\frac{x^{5}dx}{\left(e^{x}-1\right)(1-e^{-x})}-KT^{3}\label{Eq: Bloch-Gru}\end{equation}
where $x$= $\Theta_{D}/T$, $\Theta_{D}$ is the Debye temperature,
$\rho_{0}$ is the temperature independent residual resistivity and
$R\,\mathrm{and}\,~K$ are the coefficients of the phonon contribution
to the resistivity (second term) and the Mott \textit{s}-\textit{d}
inter band scattering (third term) respectively. The fit to the resistivity
curve\textcolor{black}{s} yields $\Theta_{D}=126$~K, $\rho_{0}=18\,\mu\Omega\,\mathrm{cm}$,
and $R=0.273\,\mu\Omega\,\mathrm{cm}$ for J // {[}100{]} and $\Theta_{D}=125$
K, $\rho_{0}=1.4\,\mu\Omega\,\mathrm{cm}$ and $R=0.011\,\mu\Omega\,\mathrm{cm}$
for J // {[}001{]}, the value of $K$ was found to be zero for both
the directions. The Debye temperature remains nearly same but there
is an order of decrease in the magnitude of phonon contribution to
the resistivity for J // {[}001{]} inferred from the values of R.
The Debye temperature in the Bloch-Gr$\mathrm{\ddot{u}}$neisen relation
often differs from the value of $\Theta_{D}$ obtained from heat capacity
data. In principle, the value of $\Theta_{D}$ obtained from Bloch-Gr$\mathrm{\ddot{u}}$neisen
relation should considerably differ from that obtained from heat capacity
because the former takes into account only the longitudinal phonons
\cite{Gopal}.

\subsection{CeAu$_{2}$Ge$_{2}$}

\begin{figure}
\includegraphics[width=0.5\textwidth]{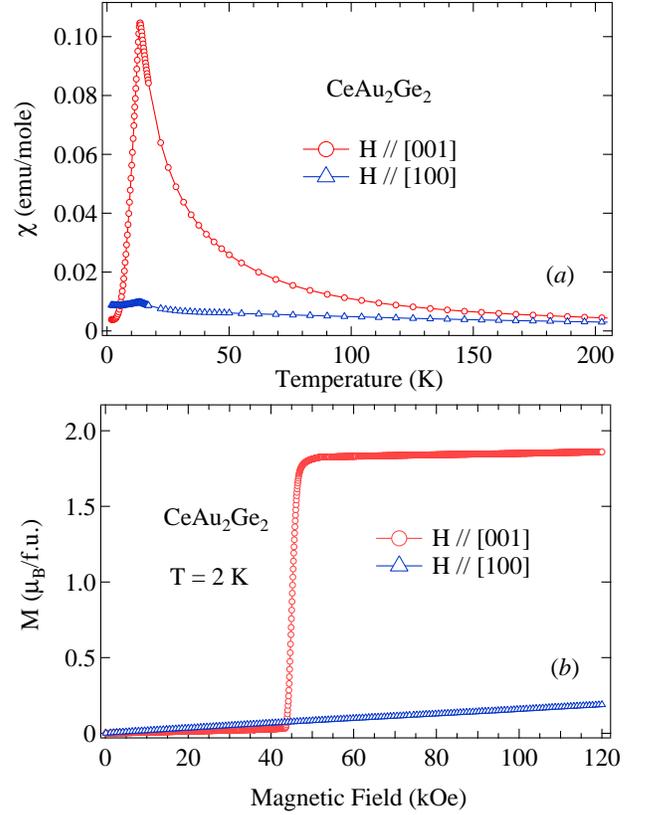}

\caption{(Color online)\label{Fig. MT-MH_Ce} a) Magnetic susceptibility of
CeAu$_{2}$Ge$_{2}$ with magnetic field (3kOe) applied along the
two crystallographic directions. b) Magnetic isotherm at 2~K for the
same with field along both the crystallographic directions.}

\end{figure}
Figure\ref{Fig. MT-MH_Ce}a shows the magnetic susceptibility of CeAu$_{2}$Ge$_{2}$
from 1.8~K to 300~K in a magnetic field of 3~kOe along the two
crystallographic directions ({[}100{]} and {[}001{]}). The susceptibility
with field parallel to {[}001{]} direction shows an antiferromagnetic
transition at T$_{{\rm N}}$~=~13.5~K, less than that observed
from neutron diffraction on a polycrystalline sample (16~K)~\cite{Loidl}.
The sharp drop of susceptibility below the N$\grave{e}$el temperature
indicates that the moments are aligned antiferromagnetically along
the {[}001{]} direction in possibly a collinear arrangement. The behavior
indicates that the {[}001{]} axis is the easy axis of magnetization
for CeAu$_{2}$Ge$_{2}$ as reported by Loidl~\textit{et al}~\cite{Loidl}.
With field along {[}100{]} direction, the susceptibility remains below
that of {[}001{]} direction in the entire temperature range followed
by a~Kink at the ordering temperature, indicating the hard axis of
magnetization. Curie-Weiss fits of the inverse susceptibility in the
paramagnetic state gives effective moment ($\mu_{{\rm eff}}$ ) and
paramagnetic Curie temperature ($\theta_{{\rm p}}$) as 2.54~$\mu_{{\rm B}}$/Ce
and 2.54~$\mu_{{\rm B}}$/Ce and -63~K and 27~K for field parallel
to {[}100{]} and {[}001{]} directions, respectively. The obtained
effective moment for both the axes is equal to the expected theoretical
value (2.54~$\mu_{{\rm B}}$/Ce). The polycrystalline
average of $\theta_{{\rm p}}$ is -34~K, which is in tune with the
antiferromagnetic nature of this compound. The large negative value
also indicates the presence of possible~Kondo interaction. The magnetic
isotherms of the compound with field along both the crystallographic
directions ({[}100{]} and {[}001{]}) are shown in Fig.~\ref{Fig. MT-MH_Ce}b.
The linear behavior of the magnetization at 2~K and up to $\approx$
43~kOe with the field along {[}001{]} axis confirms the antiferromagnetic
nature of the compound in the magnetically ordered state. The magnetization
undergoes a spin flip type metamagnetic transition at the critical
field H$_{\mathrm {C}}~\approx$~43~kOe followed by near saturation
at high fields. The saturation moment obtained at 2~K and 120~KOe
is 1.86 $\mu_{{\rm B}}$, less than the the theoretical saturation
moment of 2.14 $\mu_{{\rm B}}$ and is in agreement with the neutron
diffraction results~\cite{Loidl}. %
\begin{figure}
\includegraphics[width=0.5\textwidth]{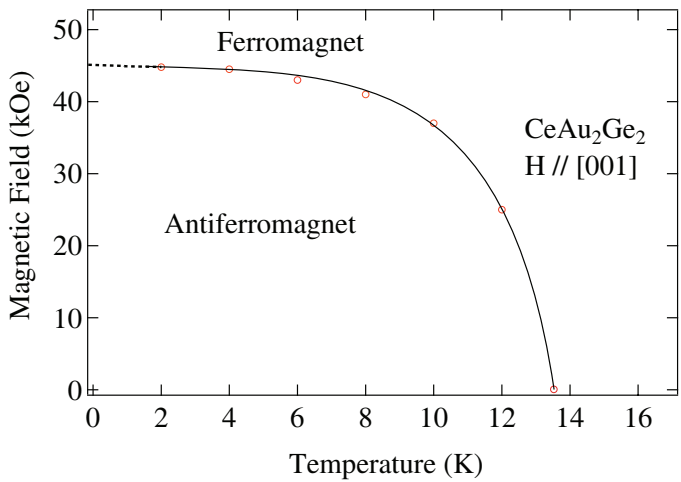}

\caption{(Color online)\label{Fig. Mag-Phase} Magnetic Phase Diagram of CeAu$_{2}$Ge$_{2}$.}

\end{figure}
From the differential plots of the isothermal magnetization curves
(not shown here), we have constructed the magnetic phase diagram as
depicted in Fig.~\ref{Fig. Mag-Phase}. The critical field H$_{\mathrm{C}}$
decreases with increase in temperature and finally vanishes at T$_{N}$.
At low temperatures and for fields less than $\approx$ 45~KOe,
the system is in a purely antiferromagnetic state as indicated and
enters into the field induced ferromagnetic state at higher fields.

\begin{figure}
\includegraphics[width=0.5\textwidth]{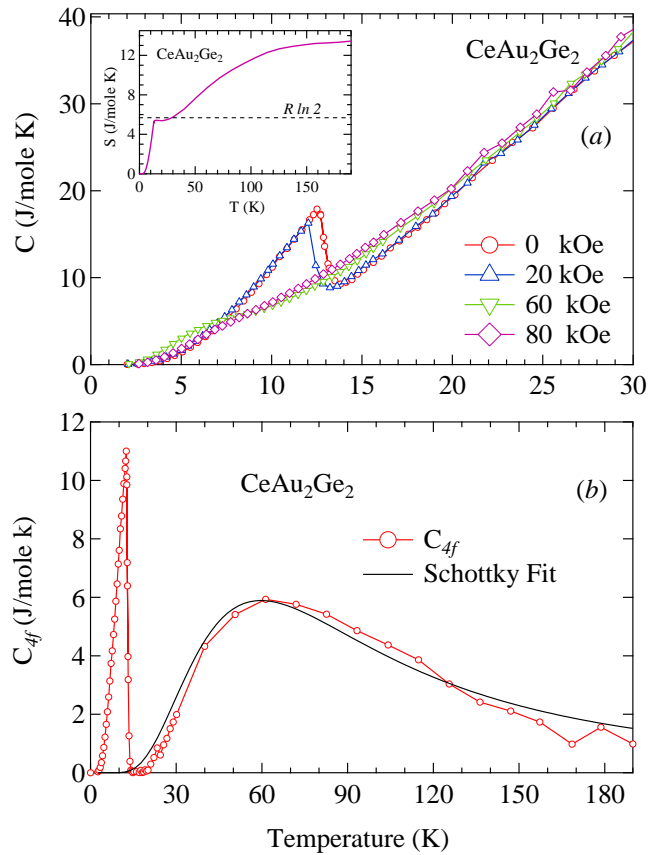}

\caption{(Color online)\label{Fig. HC_Ce} a) Heat capacity of CeAu$_{2}$Ge$_{2}$
in presence and absence of magnetic fields. The inset shows the 4\textit{f}
entropy against temperature for the same. b) Magnetic contribution
(C$_{4f}$ ) to the heat capacity of CeAu$_{2}$Ge$_{2}$ with a Schottky
fit.}

\end{figure}
The heat capacity behavior of the compound in zero and applied fields
is shown in Fig.~\ref{Fig. HC_Ce}. An anomaly in the heat capacity
at 13.5~K with a peak height of $\approx$ 11 J/mol~K confirms the
bulk magnetic ordering of Ce$^{3+}$ ions. The peak height is close
to the mean field value of 12.5 J/mol~K for spin $S=\nicefrac{1}{2}$.
The magnitude of the Sommerfield coefficient $\gamma$ was estimated
to be $\approx$ 15~mJ/mol~K$^{2}$ from the \textit{y} intercept
of the C/T vs T$^{2}$ curve (not shown). Application of the magnetic
field of 20~KOe shifts the heat capacity peak towards low temperature
and reduces its height. At higher fields (60 and 80~KOe) the peak
vanishes altogether consistent with the presence of metamagnetic transition
(43~KOe). The 4\textit{f} contribution to the heat capacity of CeAu$_{2}$Ge$_{2}$ C$_{4f}$, Fig.~\ref{Fig. HC_Ce}b
was extracted by subtracting the heat capacity of LaAu$_{2}$Ge$_{2}$.
Besides the peak at T$_{N}$, C$_{4f}$ exhibits
a broad peak centered around 60~K arising due to the Schottky contribution
from the thermal variation of the population of excited CEF levels.
The entropy calculated using the expression $S_{4f}=\int_{0}^{T}\frac{C_{4f}}{T}dT$,
plotted as a function of temperature is shown in the inset of Fig.
\ref{Fig. HC_Ce}a.The entropy
is 5.5 J/mol~K at T$_{N}$, close to the value for a well isolated
doublet ground state and attains a value of 13.4 J/mol~K at 190~K,
comparable to the theoretically expected value of \textit{Rln}\textcolor{black}{6
(14.9 J/mol~K). }

\begin{figure}
\includegraphics[width=0.5\textwidth]{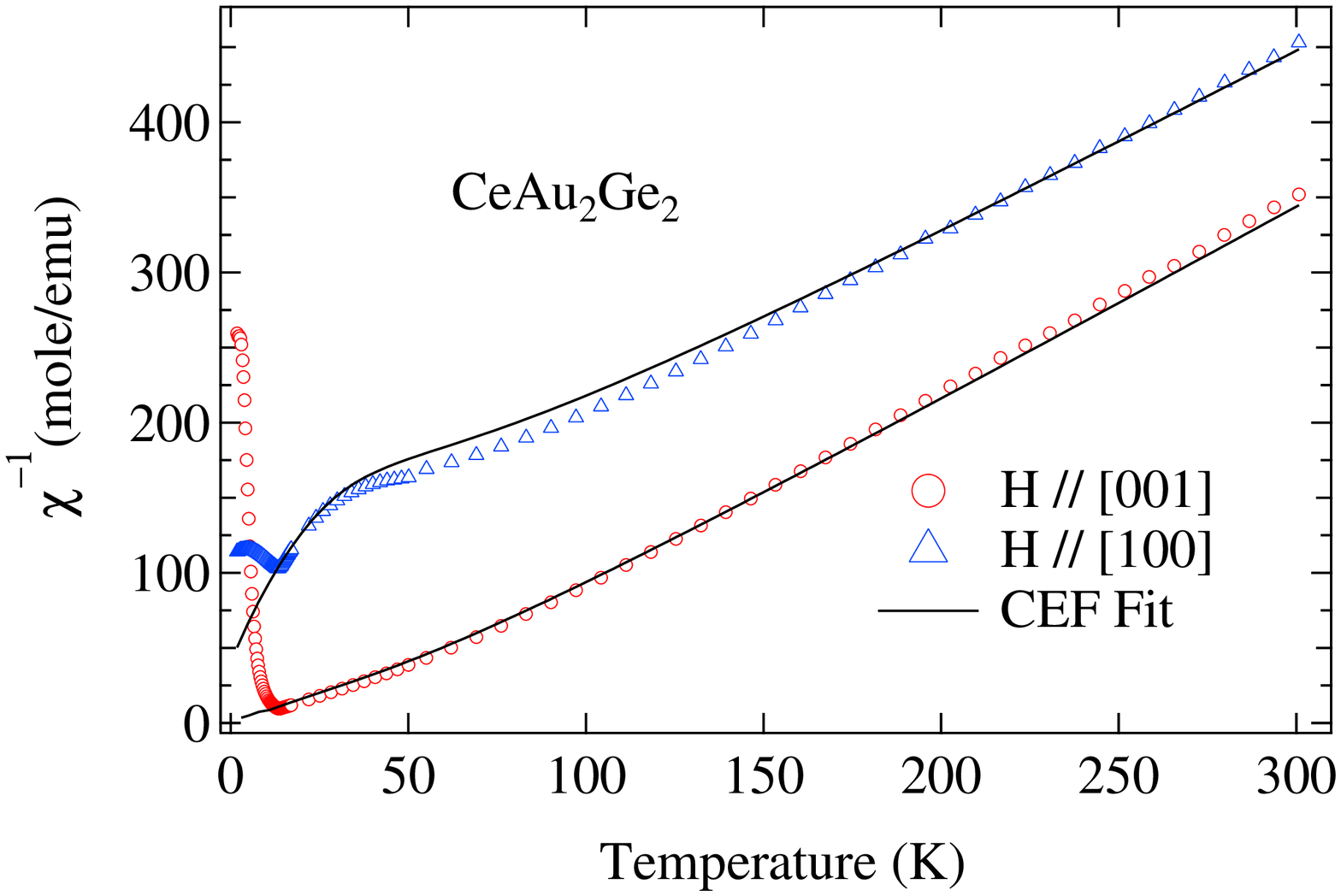}

\caption{(Color online)\label{Fig. CEF-fit} Inverse susceptibility of CeAu$_{2}$Ge$_{2}$
for H // {[}100{]} and {[}001{]} directions. The solid line through
the data points represent the crystal electric field fit.}

\end{figure}
In order to understand the magnetocrystalline anisotropy and to~Know
about the crystal field energy level splittings of the R$^{3+}$ ion
in RAu$_{2}$Ge$_{2}$, we have performed the CEF calculations using
the point charge model. The rare-earth atom in this series of compounds
occupy the $2a$ Wyckoff position which has the tetragonal point symmetry.
The CEF Hamiltonian for a tetragonal symmetry is given by, 
\begin{equation}
\mathcal{H}_{{\rm CEF}}=B_{2}^{0}O_{2}^{0}+B_{4}^{0}O_{4}^{0}+B_{4}^{4}O_{4}^{4}+B_{6}^{0}O_{6}^{0}+B_{6}^{4}O_{6}^{4},\label{Eq. CEF Hamiltonian}
\end{equation}

 where $B_{\ell}^{m}$ and $O_{\ell}^{m}$ are the CEF parameters
and the Stevens operators, respectively~\cite{Stevens,Hutchings}.
The CEF susceptibility is defined as
\begin{widetext} \begin{equation}
\chi_{{\rm CEF}i}=N(g_{J}\mu_{{\rm B}})^{2}\frac{1}{Z}\left(\sum_{m\neq n}\mid\langle m\mid J_{i}\mid n\rangle\mid^{2}\frac{1-e^{-\beta\Delta_{m,n}}}{\Delta_{m,n}}e^{-\beta E_{n}}+\sum_{n}\mid\langle n\mid J_{i}\mid n\rangle\mid^{2}\beta e^{-\beta E_{n}}\right),\label{Eq. Chi_CEF}\end{equation}
 \end{widetext} where $g_{J}$ is the Landé $g$\,-\,factor, $E_{n}$
and $\mid\! n\rangle$ are the $n$th eigenvalue and eigenfunction,
respectively. $J_{i}$ ($i$\,=\,$x$, $y$ and $z$) is the component
of the angular momentum, and $\Delta_{m,n}\,=\, E_{n}\,-\, E_{m}$,
$Z\,=\,\sum_{n}e^{-\beta E_{n}}$ and $\beta\,=\,1/k_{{\rm B}}T$.
The magnetic susceptibility including the molecular field contribution
$\lambda_{i}$ is given by \begin{equation}
\chi_{i}^{-1}=\chi_{{\rm CEF}i}^{-1}-\lambda_{i}.\label{Eq. Chi - Lambda}\end{equation}
For Ce$^{3+}$ ions the $O_{6}$ terms in the above Hamiltonian vanishes
resulting in only three crystal filed parameters. The inverse susceptibility
of CeAu$_{2}$Ge$_{2}$ with field along both the crystallographic
directions was fitted to the mentioned CEF model as shown in Fig.
\ref{Fig. CEF-fit}. The CEF parameters obtained for the best fit
are $B_{2}^{0}$ = -6.4~K, $B_{4}^{0}$ = -0.27~K and $B_{4}^{4}$=
2.6~K with a molecular field contribution of $\lambda(100)$ = -41
K and $\lambda(001)$ = -8~K respectively for field along {[}100{]}
and {[}001{]} directions. The negative value of $\lambda$ supports
the antiferromagnetic exchange interaction among the moments. The
susceptibility could be fitted using a set of values of CEF parameters
but only those values were considered which could also fit the experimentally
obtained Schottky anomaly. The CEF split energy levels obtained from
the above CEF parameters are three doublets $\triangle_{0}$ =
0~K, $\triangle_{1}$= 128~K and $\triangle_{2}$= 199~K which
are in excellent agreement with the values derived from neutron scattering
results on polycrystalline sample. These energy
levels were used to calculate the Schottky contribution using the
equation\begin{equation}
C_{Sch}\left(T\right)=R\left[\frac{\sum_{i}g_{i}e^{-E_{i}/T}\sum_{i}g_{i}E_{i}^{2}e^{-E_{i}/T}-\left[\sum_{i}g_{i}E_{i}e^{-E_{i}/T}\right]^{2}}{T^{2}\left[\sum_{i}g_{i}e^{-E_{i}/T}\right]^{2}}\right]\label{Eq: Schottky}\end{equation}
where $R$ is a gas constant, $E_{i}$ is the energy in units of temperature
and $g_{i}$ is the degeneracy of the energy level. The calculated
Schottky heat capacity is in good agreement with the observed one
as seen in Fig.~\ref{Fig. HC_Ce}b. According to the mean field theory,
the CEF parameter $B_{2}^{0}$ is related to the exchange constant
and paramagnetic Curie temperature by the relation~\cite{Jensen}

\begin{equation}
\theta_{{\rm p}}^{[001]}=\frac{J(J+1)}{3k_{{\rm B}}}\mathcal{J_{{\rm ex}}}^{[001]}-\frac{(2J-1)(2J+3)}{5k_{{\rm B}}}B_{2}^{0},\label{Eq. Jex001}\end{equation}
 \begin{equation}
\theta_{{\rm p}}^{[100]}=\frac{J(J+1)}{3k_{{\rm B}}}\mathcal{J_{{\rm ex}}}^{[100]}+\frac{(2J-1)(2J+3)}{10k_{{\rm B}}}B_{2}^{0}.\label{Eq. Jex100}\end{equation}

Substituting the values of the parameters we obtain $\mathcal{J}_{{\rm ex}}^{[100]}$
= -14.5~K and $\mathcal{J}_{{\rm ex}}^{[001]}$ = -4.7~K. The negative
value of the exchange constant along both the principal directions
imply an overall antiferromagnetic interaction among the moments.

\begin{figure}
\includegraphics[width=0.5\textwidth]{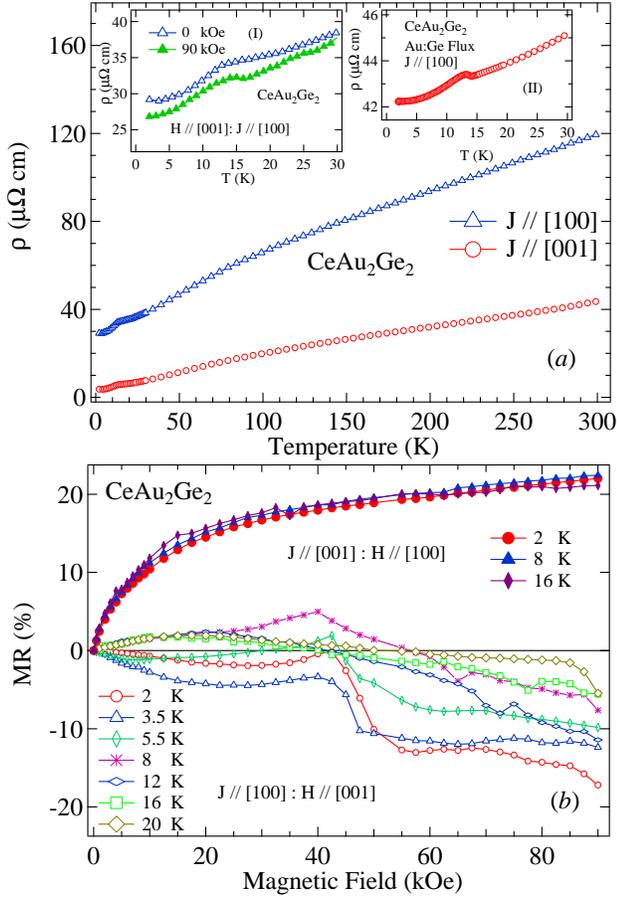}

\caption{(Color online)\label{Fig. Ce-Rho} a) Resistivity of CeAu$_{2}$Ge$_{2}$
with J // {[}100{]} and {[}001{]} direction. The inset (I) shows the
low temperature portion with J // {[}100{]} and H // {[}001{]} direction.
The inset (II) shows the resistivity of CeAu$_{2}$Ge$_{2}$ single
crystal grown with Au-Ge flux (J // {[}100{]}) for comparison with
the presently studied Bi flux grown single crystal. b) The transverse
magnetoresistance of the same with J // {[}100{]}, H // {[}001{]}
and J // {[}001{]}, H // {[}100{]} at various temperatures are shown.}

\end{figure}
Fig.~\ref{Fig. Ce-Rho}a shows the temperature dependence of resistivity
for CeAu$_{2}$Ge$_{2}$ with current parallel to {[}100{]} and {[}001{]} directions. The resistivity
shows a metallic behavior down to $\approx$ 130~K followed by a broad
hump and then drops faster at the ordering temperature ( T$_{N}$
= 13.5~K) due to the gradual freezing of the spin disorder resistivity.
The broad hump below $\approx$ 130~K is due to the CEF effects. The
thermally induced variation of the fractional Boltzmann occupation
of the CEF levels changes the otherwise constant spin disorder resistivity
and is qualitatively in agreement with the calculated CEF split energy
levels with second excited state lying at 199~K. Similar
to LaAu$_{2}$Ge$_{2}$ the resistivity with current
along {[}100{]} direction is higher than along
{[}001{]} indicating the structural anisotropy in this compound. The
inset (I) of Fig.~\ref{Fig. Ce-Rho}a shows the low temperature resistivity
in zero and applied field of 90~KOe for J // {[}100{]} and H // {[}001{]}.
The resistivity curve drops at the ordering temperature of the compound
as usual. Application of magnetic field (90~KOe) reduces the resistivity
with appearance of a hump just above the ordering temperature. The
negative magnetoresistance is in agreement with the (field induced)
ferromagnetic behavior of the compound at 90~KOe. The negative magnetoresistance
above the ordering temperature is due to the reduction in the spin
disorder scattering in presence of field (90~KOe) compared to the
zero field one. The appearance of hump just above the ordering temperature
with application of field is an anomalous behavior. The exact reason
for the observed behavior is not~Known but one of the possibilities
is discussed below. At 90~KOe and low temperatures
the compound is in the field induced ferromagnetic state, increase
in temperature may result in the formation of short range spin fluctuations
as a consequence of competition between the thermal energy and the
external field causing the scattering of the conduction electrons.

The inset (II) of Fig.~\ref{Fig. Ce-Rho}a shows
the low temperature part of the resistivity (J~//~{[}100{]}) of CeAu$_{2}$Ge$_{2}$
single crystal grown using Au:Ge flux. Unlike the resistivity of Bi
flux grown single crystal, the resistivity of Au:Ge flux grown crystal
shows a peak at the ordering temperature T$_{\mathrm{N}}$ = 13.5~K. This type of peak at the ordering temperature is generally attributed
to the superzone gap effect. But considering the previous neutron
diffraction studies which shows that CeAu$_{2}$Ge$_{2}$ is a simple
collinear antiferromagnet \cite{Loidl} with
a propagation vector (0, 0, 1), where such type of superzone gap
effect are not usually observed, we claim that the single crystal
grown using Bi-flux is good compared to the Au:Ge flux grown single
crystals. Although, the other preliminary magnetic property of Au:Ge
flux grown single crystal shows similar behavior (not shown here for
brevity) to that of Bi flux grown single crystals. 

\begin{figure}
\includegraphics[width=0.5\textwidth]{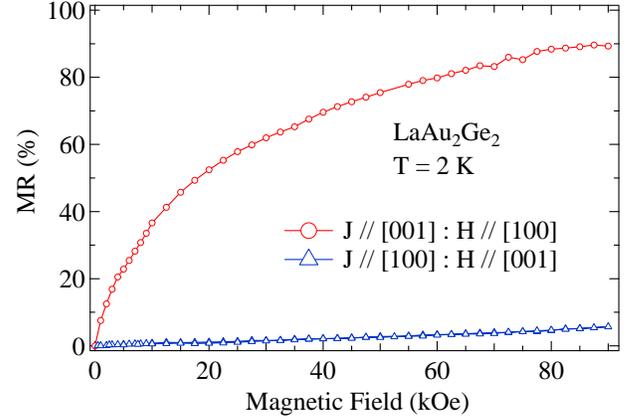}

\caption{(Color online)\label{Fig. La_MR} Transverse Magnetoresistance of
LaAu$_{2}$Ge$_{2}$ at 2~K. }

\end{figure}
The transverse MR resistance of CeAu$_{2}$Ge$_{2}$
as a function of field and temperatures is shown in (Fig.~\ref{Fig. Ce-Rho}b).
The magnetoresistance of the compound was calculated using the relation
MR = {[}R(H)-R(0){]}/R(0) (MR~\%~=~MR~*~100). Before discussing
the MR behavior of CeAu$_{2}$Ge$_{2}$ we first
discuss the MR behavior of LaAu$_{2}$Ge$_{2}$
(Fig.~\ref{Fig. La_MR}) which would be helpful in understanding the
MR behavior of CeAu$_{2}$Ge$_{2}$. The MR of
LaAu$_{2}$Ge$_{2}$ with J // {[}100{]} and H
// {[}001{]} (Fig.~\ref{Fig. La_MR}) increases with field up to $\approx$~5~\% at 90~KOe. Whereas with J // {[}001{]} and H // {[}100{]} the MR
increases anomalously to $\approx$ 90 \% at 90~KOe. The large anisotropic
behavior in the MR of the nonmagnetic compound indicates the presence
of significant anisotropy in the Fermi surface of the compound. Referring
to the MR of CeAu$_{2}$Ge$_{2}$ at 2~K and with J // {[}100{]}
and H // {[}001{]}, the curve initially decreases
with field followed by an upward turn above 30~KOe and then decreases
sharply above $\approx$ 42~KOe. The initial decrease in MR at low
fields (below 30~KOe) is surprising since the compound is in the antiferromagnetically
ordered state and a positive magnetoresistance is generally expected.
The negative magnetoresistance appears in the compounds having ferromagnetic
ordering or~Kondo behavior. The initial negative increase of MR at
low fields indicates the weak~Kondo interaction in the compound, but
sufficient enough to overcome the small positive contribution arising
from Lorentz effect and antiferromagnetic ordering. The
presence of weak~Kondo interaction is supported by the large negative
value of $\theta_{{\rm p}}$, low saturation moment of 1.86 $\mu_{B}$/Ce
in the magnetic isotherm at 2~K and a heat capacity jump of 11 J/mol
K compared to 12.5 J/mol~K. The upward turn in the MR between 30
and 42~KOe is due to the positive contribution arising from antiferromagnetic
ordering. The sharp drop in the magnetoresistance above 42~KOe corresponds
to the field induced ferromagnetic behavior of the compound, consistent
with the metamagnetic transition in the magnetic isotherm. With increase
in temperature to 3.3~K the MR at low fields become more negative
compared to that at 2~K. The behavior is also in agreement with the
Kondo behavior which exhibit a minimum in temperature variation of
MR~\cite{Zlatic}. At high temperature (above 5.5~K) the MR increases
with field due to the increasing positive contribution from antiferromagnetic
coupling and decreasing contribution from~Kondo effect. The maximum
occurs at 8~K and then decreases with temperature as expected. The
negative contribution in the paramagnetic state at high fields is
due to the decrease in the spin disorder scattering. The MR with J
// {[}001{]} and H // {[}100{]} is in sharp contrast to the former
one showing a positive contribution at all fields and nearly temperature
independent behavior. The temperature independent behavior of the
MR indicates the absence of magnetic contribution. Comparing the MR
behavior with that of the La counterpart and assuming that a similar
applies here, it is possible that an anisotropy in the Fermi surface
cause a comparatively large MR with J // {[}001{]} suppressing the
other smaller contribution. With J // {[}100{]}, because of the smaller
contribution from the Fermi surface, the other magnetic contributions
are dominant.

\subsection{PrAu$_{2}$Ge$_{2}$}

\begin{figure}
\includegraphics[width=0.5\textwidth]{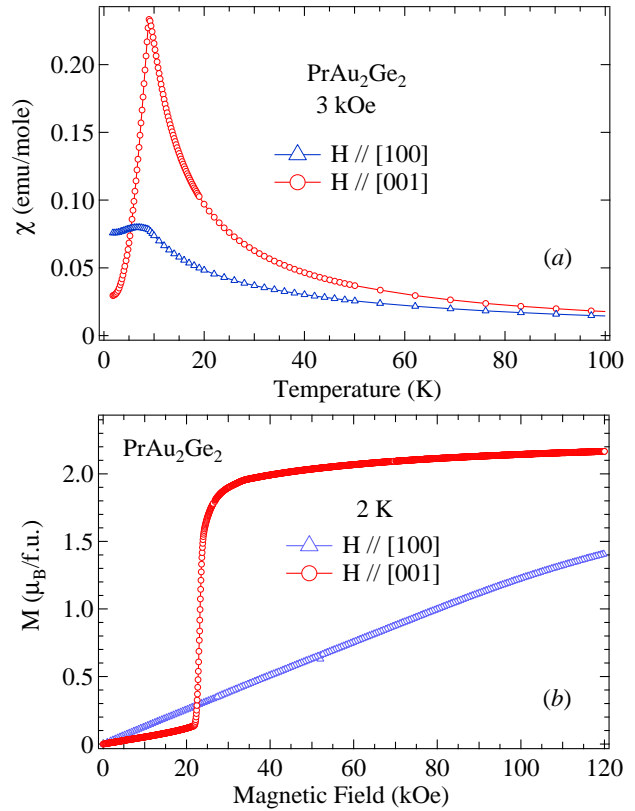}

\caption{(Color online)\label{Fig. MT-MH_Pr} a) Magnetic susceptibility of
PrAu$_{2}$Ge$_{2}$ with magnetic field (3kOe) applied along the
two crystallographic directions. b) Magnetic isotherm at 2~K for the
same with field along both the crystallographic directions.}

\end{figure}
PrAu$_{2}$Ge$_{2}$ orders antiferromagnetically at 9~K with {[}001{]}
as the easy axis of magnetization (Fig.~\ref{Fig. MT-MH_Pr}a) similar
to CeAu$_{2}$Ge$_{2}$. In the paramagnetic state, the magnetic susceptibility
was fitted to the Curie-Weiss law. \textcolor{black}{The fit gives
$\mu_{{\rm eff}}$ and $\theta_{{\rm p}}$ as 3.57~$\mu_{{\rm B}}$/Pr
and -10~K and 10~K for field parallel to {[}100{]} and {[}001{]}
directions, respectively.} The value of $\mu_{{\rm eff}}$ is equal
to the theoretically expected value of Pr$^{3+}$ ion. The magnetic
isotherm at 2~K (Fig.~\ref{Fig. MT-MH_Pr}b) with H // {[}001{]} shows
a linear behavior up to 20~KOe in confirmation with the antiferromagnetic
nature of the compound. At $\approx$22~KOe (H // {[}001{]}) the compound
undergoes a spin flip type metamagnetic transition followed by slow
increase with field attaining a magnetic moment of $\approx$ 2.2
$\mu_{B}$/f.u. at 120~KOe. The moment is less than the saturation
moment of Pr$^{3+}$ ion. The less moment may be due to the crystal
field effect. %
\begin{figure}
\includegraphics[width=0.5\textwidth]{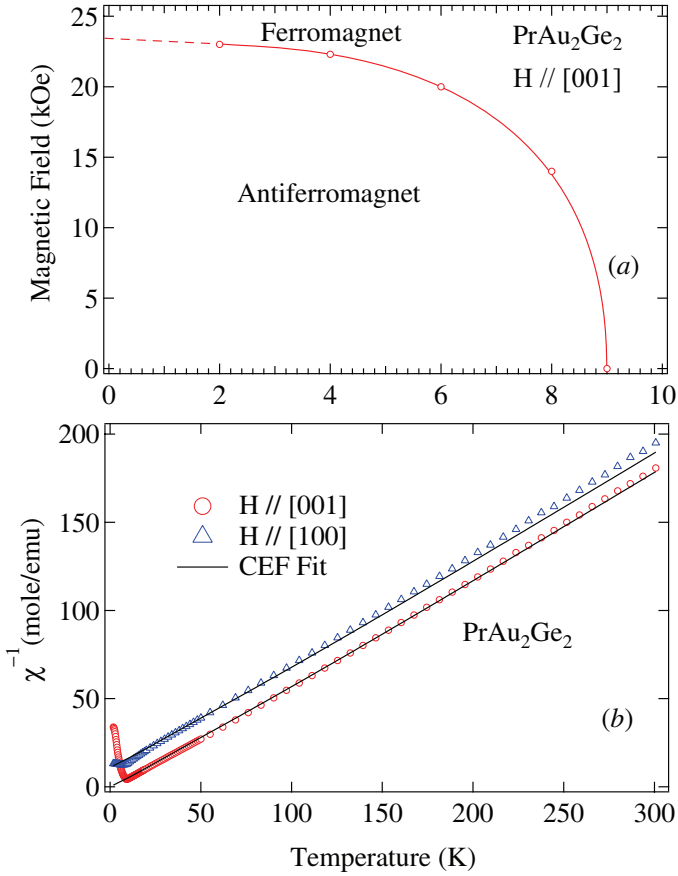}

\caption{(Color online)\label{Fig. Mag-Phase_Pr} a) Magnetic Phase Diagram
of PrAu$_{2}$Ge$_{2}$ constructed from the magnetic isotherms at
various temperatures. b) Inverse susceptibility of PrAu$_{2}$Ge$_{2}$
with crystal electric field fit.}

\end{figure}
The magnetic isotherm with field along {[}100{]} direction increases
linearly with a magnetization of $\approx$ 1.4~$\mu_{B}$/f.u. at
120~KOe, indicating the hard axis of magnetization. The magnetic phase
diagram constructed as discussed before (Fig.~\ref{Fig. Mag-Phase_Pr}a)
shows the antiferromagnetic and the field induced ferromagnetic behavior
of the compound. For detailed investigation a crystal field analysis
of the compound was done by fitting
the inverse susceptibility as shown in Fig.~\ref{Fig. Mag-Phase_Pr}b.
The Obtained value of the crystal field parameters are $B_{2}^{0}$
= -1.2~K, $B_{4}^{0}$ = 0.08~K, $B_{4}^{4}$= 0.25~K, $B_{6}^{0}$
= -0.0001K and $B_{6}^{4}$ = 0.006~K with a molecular field contribution
of $\lambda(100)$ = -10~K and $\lambda(001)$ = 0~K respectively
for field along {[}100{]} and {[}001{]} directions. Only those value
of crystal field parameters are considered which fits the susceptibility
as well as provide a satisfactory representation of the Schottky anomaly
inferred from the heat capacity measurements (described below). The
crystal field split energy levels calculated using the above crystal
field parameters are $\triangle_{0}$= 0~K (doublet), $\triangle_{1}$=
39~K, $\triangle_{2}$= 98~K, $\triangle_{3}$= 115~K, $\triangle_{4}$=
148~K, $\triangle_{5}$= 186~K(doublet) and $\triangle_{6}$=
261~K. The energy level scheme shows a doublet ground state for the
Pr$^{3+}$ ion. The exchange interaction constant obtained using Eq.
\ref{Eq. Jex100} and~\ref{Eq. Jex001} are $\mathcal{J}_{{\rm ex}}^{[100]}$
= -0.114~K and $\mathcal{J}_{{\rm ex}}^{[001]}$ = -1.27~K. The negative
value of the exchange interaction constant along both the direction
indicates an overall antiferromagnetic interaction among the moments

\begin{figure}
\includegraphics[width=0.5\textwidth]{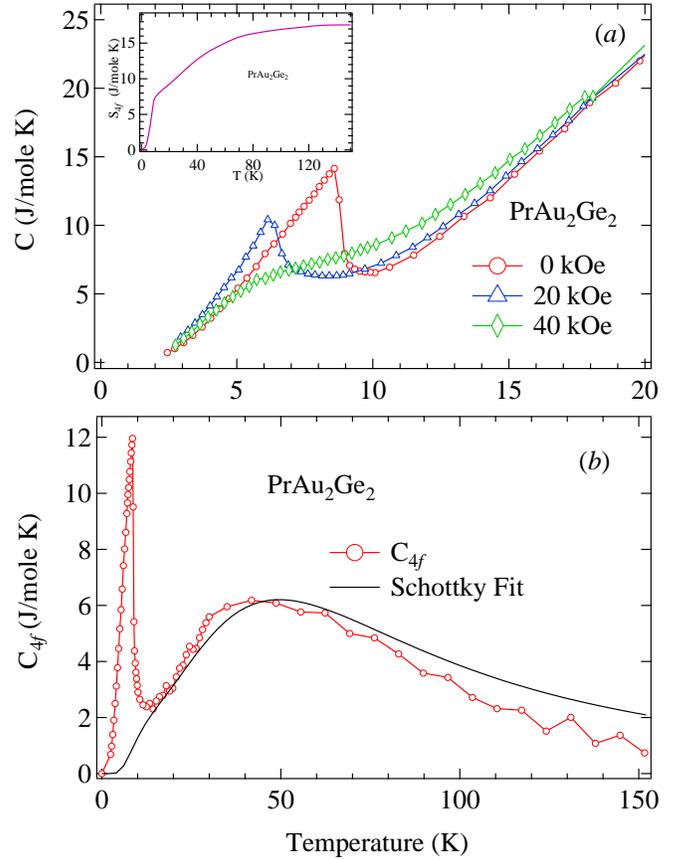}

\caption{(Color online)\label{Fig. HC_Pr} a) Heat capacity of PrAu$_{2}$Ge$_{2}$
in \textcolor{black}{0, 20 and 40~KOe.} The inset shows the 4\textit{f}
entropy against \textcolor{black}{temperature. b) }Magnetic contribution
(C$_{4f}$ ) to the heat capacity of PrAu$_{2}$Ge\textcolor{black}{$_{2}$;
the solid line shows the Schottky heat capacity calculated from the
CEF levels.}}

\end{figure}
 The heat capacity behavior of PrAu$_{2}$Ge$_{2}$
in 0, 20 and 40~KOe is shown in Fig.~\ref{Fig. HC_Pr}a. The heat
capacity shows an anomaly at the antiferromagnetic ordering temperature
of the compound. An external field of 20~KOe shifts the peak towards
low temperature as expected for an antiferromagnetically ordered compound.
At a higher field of 40~KOe, above the metamagnetic
transition field, the peak vanishes and there is a broad hump. The
magnetic contribution to the heat capacity (Fig.~\ref{Fig. HC_Pr}b)
was isolated by subtracting the heat capacity of LaAu$_{2}$Ge$_{2}$.
It shows a sharp peak at the ordering temperature followed by a Schottky
anomaly at high temperature. The entropy calculated
using the equation as mentioned before is plotted as a function of
temperature in the inset of Fig.~\ref{Fig. HC_Pr}a. It attains a
value of $\approx$~7~J/mol~K at the ordering temperature,
which exceeds substantially the entropy for a doublet ground state
with effective J = 1/2 (5.76~J/mol~K). The excess entropy appears
because of a substantial contribution to C$_{4f}$ \ due to the Schottky
heat capacity at low temperatures arising from the first excited CEF
level lying at 39~K. The total entropy obtained at 150~K is 17.6~J/mol
K, close to the expected value of 18.2 J /mol~K. The solid line in
Fig.~\ref{Fig. HC_Pr}b shows the Schottky heat capacity calculated
from the CEF split energy levels using expression~\ref{Eq: Schottky}.
An overall good agreement with the experimental C$_{4f}$ increases
the confidence in the proposed CEF level scheme. 

\begin{figure}
\includegraphics[width=0.5\textwidth]{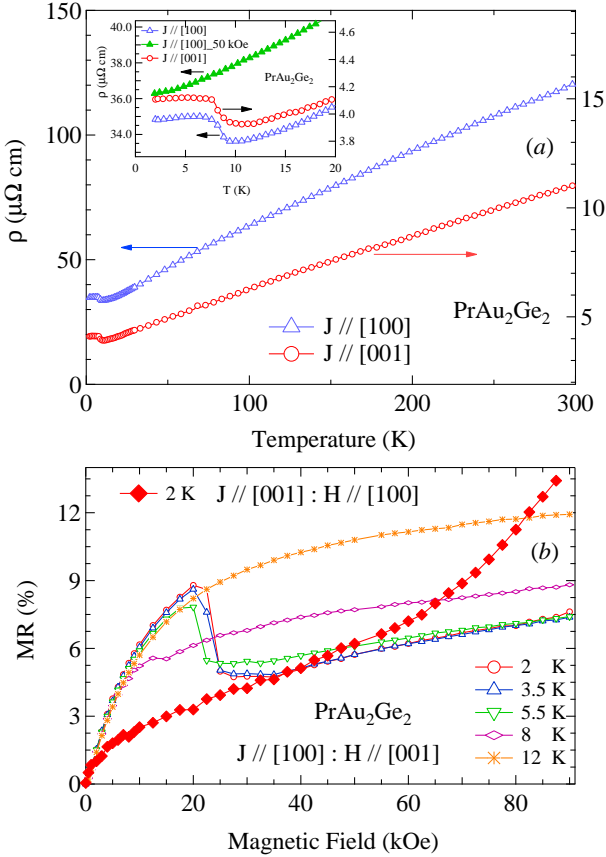}

\caption{(Color online)\label{Fig. Rho_Pr} a) Resistivity of PrAu$_{2}$Ge$_{2}$
with J // {[}100{]} and {[}001{]} direction. The inset shows the expanded
low temperature portion with arrows indicating the respective axis.
b) The transverse magnetoresistance of the same with J // {[}100{]}:
H // {[}001{]} at various temperatures and J // {[}001{]}: H // {[}100{]}
are shown.}

\end{figure}
The resistivity of PrAu$_{2}$Ge$_{2}$ with current along the two
principle crystallographic directions is shown in Fig.~\ref{Fig. Rho_Pr}a.
The resistivity with J // {[}100{]} is higher than that with J //
{[}001{]} as seen for both La and Ce compounds. The resistivity with
current along both the directions show a metallic behavior down to
the ordering temperature followed by a sharp increase at the ordering
temperature (inset of Fig.~\ref{Fig. Rho_Pr}a.) and then falls down
marginally at low temperatures. The behavior is anomalous to the one
expected below the ordering temperature. The rise of the resistivity
at the ordering temperature is attributed to the formation of energy
gap in conduction band due to the difference in the periodicity of
the antiferromagnetic configuration and lattice, also~Known as superzone
gap effect~\cite{Elliot}. The behavior indicates that moments have
antiferromagnetic interaction along both the directions with different
periodicity. Since the resistivity with J // {[}100{]} is higher than
that with J // {[}001{]} and also the rise in the resistivity at T$_{N}$
is $\approx$ 1.5~$\mu\Omega$ cm for the former case and $\approx$
0.3 $\mu\Omega$ cm for the later, we infer a higher energy gap along
the {[}100{]} direction compared to that along {[}001{]} direction.
The superzone gap effect in few cases affects the
Fermi surface and reduces the density of states at the Fermi level
resulting in the marginal decrease of the resistivity below $\mathrm{T}\mathrm{_{N}}$.
In an applied field of 50~KOe (field well above the metamagnetic transition)
the upturn in the resistivity at T$_{N}$ vanishes due to the disappearance
of superzone gap effect and the resistivity decreases linearly for
both the direction (shown only for J // {[}100{]}: H // {[}001{]})
of current. 

The transverse MR of the compound at various temperatures is shown
in Fig.~\ref{Fig. Rho_Pr}b. The magnetoresistance at 2~K and with
J // {[}100{]} and H // {[}001{]} initially increases with field up
to $\approx$ 20~KOe followed by a sharp drop and then increases at
higher field. The sharp drop in MR is due to the field induced ferromagnetic
behavior of the compound. The sharp drop moves towards
lower field with increase in temperature as expected and vanishes
in the paramagnetic state (12~K). The initial increase of the positive
MR with field (below 20~KOe) reflects contributions from both the
fluctuations induced in the antiferromagnetic state by the applied
field and positive Lorentz contribution from the conduction electrons.
The latter is also likely responsible for the increase of MR at higher
fields and low temperatures. Its presence is also evident from the
MR at 12~K (paramagnetic state) which increases and saturates at high
fields. With J // {[}001{]} and H // {[}100{]} the MR increases with
field followed by an increase in the slope at higher fields (above
60~KOe). The increase in the slope at high fields may be due to the
gradual orientation of the ordered moments towards {[}100{]} direction.

\section{Conclusion}

In conclusion, we have studied the magnetic behavior of the single
crystalline RAu$_{2}$Ge$_{2}$ (R La, Ce and Pr) compounds. LaAu$_{2}$Ge$_{2}$
shows a Pauli-paramagnetic behavior with both resistivity and heat
capacity measurements reflecting its nonmagnetic behavior. The heat
capacity shows a presence of both optical and acoustic modes of vibration
of its atoms. The magnetoresistance behavior indicates a large anisotropy
in the geometry of its Fermi surface. CeAu$_{2}$Ge$_{2}$ shows a
simple antiferromagnetic behavior with T$_{N}$ = 13.5~K and {[}001{]}
as the easy axis of magnetization. The compound undergoes a spin flip
like metamagnetic transition at 2~K and at a critical field H$_{\mathrm{C}}$
$\approx$~43~KOe driving the compound to a field induced ferromagnetic
state. The low temperature behavior of the compound
indicates a presence of weak~Kondo interaction. PrAu$_{2}$Ge$_{2}$
orders antiferromagnetically at T$_{N}$ = 9~K with {[}001{]} as the
easy axis of magnetization. This compound also undergoes a spin flip
type metamagnetic transition at 2~K and at a critical field H$_{\mathrm{C}}$
$\approx$ 22~KOe. Crystal electric field analysis for CeAu$_{2}$Ge$_{2}$
and PrAu$_{2}$Ge$_{2}$ shows a doublet ground state for both the
compounds.


\begin{thebibliography}{26}
\bibitem{Franz}W. Franz, A. GrieBel, F. Steglich, and D. Wohlleben,
Z. Physik B \textbf{31}, 7 (1978).

\bibitem{Knopp}G.~Knopp, A. Loidl,~K.~Knorr, L. Pawlak, M. Duczmal,
R. Caspary, U. Gottwick, H. Spille, F. Steglich, and A.P. Murani,
Z. Phys. B - Condensed Matter \textbf{77}, 95 (1989) . 

\bibitem{Yoshida}J. Yoshida, S. Abe, D. Takahashi, Y. Segawa, Y.
Komai, H. Tsujii,~K. Matsumoto, H. Suzuki and Y. Onuki, Phys. Rev.
Lett. \textbf{101}, 256402 (2008).

\bibitem{Mathur}N. D. Mathur, F. M. Grosche, S. R. Julian, I. R.
Walker, D. M. Freye, R.~K. W. Haseiwimmer and G. G. Lomzarich, Nature
(London), \textbf{394}, 39 (1998).

\bibitem{Vargoz}E. Vargoz and D. Jaccard, J. Magn. Magn. Mater.,
\textbf{177}, 294 (1998).

\bibitem{Movshovich}R. Movshovich, T. Graf, D. Mandrus, J. D. Thompson,
J. L. Smith, and Z. Fisk, Phys. Rev. B \textbf{53},8241 (1996).

\bibitem{Thamizh}A. Thamizhavel, R.~Kulkarni, and S.~K. Dhar, Phys.
Rev. B \textbf{75}, 144426 (2007).

\bibitem{Mulders}A. M. Mulders, A. Yaouanc, P. Dalmas de R$\acute{e}$otier,
P. C. M. Gubbens, A. A. Moolenaar, B. Fak, E. Ressouche,~K. Proke$\breve{s}$,
A. A. Menovsky, and~K. H. J. Buschow, Phys. Rev. B \textbf{56}, 8752
(1997).

\bibitem{Vejpravova}J. Vejpravova, J. Prokleska, V. Sechovsky, J.
Magn. Magn. Mater., \textbf{316}, e374 (2007).

\bibitem{Szytula}A. Szytula, J. Leciejewich, H. Binzycka, Phys. Stat.
Sol. A \textbf{58, }67 (1980).

\bibitem{Kawae}T.~Kawae, M. Mito, M. Hitaka, F. Ichikawa, T. Shigeoka,
N. Iwata and~K. Takeda, J. Phys. Soc. Jpn. \textbf{69}, 586 (2000).

\bibitem{Blanco}J. A. Blanco, R. M. Nicklow and D. Schmitt. Physica
B \textbf{213}, 327 (1995).

\bibitem{Szytula-1}E. Wawrzy$\acute{n}$ska ,M. Ba\l{}anda ,S. Baran,
J. Leciejewicz, B. Penc, N. St$\ddot{\mathrm{u}}$ßer and A. Szytu\l{}a,
J. Phys.:Condens. Matter \textbf{17} 1037 (2005).

\bibitem{Hiebl}K. Hiebl, C. Horvath, P. Rogl, M. J. Sienko, Solid
State Comm. \textbf{48}, 211 (1983).

\bibitem{Loidl}A. Loidl,~K.~Knorr, G.~Knopp, A.~Krimmel, R. Caspary,
A. B$\ddot{\mathrm{o}}$hm, G. Sparn, C. Geibel, F. Steglich, A. P.
Murani, Phys. Rev. B \textbf{46}, 9341 (1992).

\bibitem{Nishimura}K. Nishimura, M. Yamatoto,~K. Mori, J. Magn. Magn.
Mater., \textbf{177-181, }1087-i088 (1998). 

\bibitem{Rodrigues}J. Rodrigues-Carvajal, Physica B (Amsterdam) \textbf{192,
}55 (1992).

\bibitem{Boer}FR de Boer, J. C. P.~Klaasse, P. A. Veenhuizen, A.
Bohm, C. D. Bredl, U. Gottwick, H. M. Mayer, L. Pawlak, U. Rauchschwalbe,
H. Spille and F. Steglich, J. Magn. Magn. Mater., \textbf{63-64, }91-94
(1987). 

\bibitem{Lawless}W. N. Lawless, Phys. Rev. B \textbf{14}, 134 (1976).

\bibitem{Chambers}R. G. Chambers, Proc. Phys. Soc. London \textbf{78},
941 (1961).

\bibitem{Gopal}E. S. R. Gopal, Specific Heat at Low Temperatures,
(Plenum Press), \textbf{Chap. 2} (1966) and referances theirin.

\bibitem{Stevens}~K. W. H. Stevens, Proc. Phys. Soc., London, Sect.
A\textbf{65}, 209 (1952).

\bibitem{Hutchings} M. T. Hutchings, in \textit{Solid State Physics:
Advances in Research and Applications}, edited by F. Seitz and B.
Turnbull (Academic, New York, 1965), Vol.16, p.227.

\bibitem{Jensen}J. Jensen and A. R. Mackintosh, Rare earth magnetism
structures and excitations, Calrendon press, Oxford \textbf{Chap 2},
p. 73 (1991).

\bibitem{Zlatic}V Zlatic, J. Phys. F: Met. Phys. \textbf{\textcolor{black}{11}},
2147 (1981)

\bibitem{Elliot}R. J. Elliot and F. A. Wedgwood: Proc. Phys. Soc.
\textbf{81, }846 (1963).
\end{thebibliography}
\end{document}